\documentclass[aip,rsi,reprint]{revtex4-1}
\usepackage{amsmath}
\usepackage{amsfonts}
\usepackage{amssymb}
\usepackage{graphicx}
\usepackage{array}

\begin{document}

\title{A Transimpedance Amplifier for Remotely Located Quartz Tuning Forks}
\author{Ethan Kleinbaum$^1$ and G\' abor A. Cs\'{a}thy$^{1,2}$}
\affiliation{${}^1$Department of Physics, Purdue University, West Lafayette, IN 47907, USA\\
${}^2$Birck Nanotechnology Center, Purdue University, West Lafayette, IN 47907, USA}
\date{\today}

\begin{abstract}
The cable capacitance in cryogenic and high vacuum applications of quartz tuning forks 
imposes severe constraints on the bandwidth and noise performance of the measurement.
We present a single stage low noise transimpedance amplifier with a bandwidth exceeding 1~MHz
and provide an in-depth analysis of the dependence of the amplifier parameters on the cable capacitance. 
\end{abstract}
\maketitle

Quartz tuning forks (QTF) have proven to be a versatile tool in optical near-field microscopy \cite{karrai},
in scanning probe microscopy \cite{intro:afm}, 
in studies of the quantum liquids \cite{intro:He3}, in mass sensing in biological systems \cite{intro:biosensor}, 
in magnetometry \cite{magn}, and in low temperature
thermometry \cite{intro:thermometry}. Their popularity stems from their high quality factor
and from the possibility of electrical excitation and detection. In a growing number of 
cryogenic and high vacuum applications the QTF cannot be located close the the measurement electronics and,
therefore, cables with significant capacitances are often used to connect them
\cite{intro:thermometry,coldq1,coldq2,coldq3,coldq4,coldq5,coldq6,coldq7,uhv}.
 
QTFs operate in the 20-150~kHz frequency range but novel devices, called needle oscillators, may resonate at frequencies as high as 1~MHz
\cite{heike,app:extension650kHz,app:extension1MHz}. At such frequencies and in the presence of considerable cable capacitances the preferred
measurement circuits employ current detection with a transimpedance amplifier (TIA).
Many applications require small amplitude oscillations and, therefore, benefit from low noise amplification. 

Large cable capacitances have a negative impact on the bandwidth and noise performance of TIAs.
For low temperature QTF applications cryogenic TIAs were developed which are placed in close proximity to the QTF and, therefore,
reduce the cable capacitance \cite{app:extension1MHz,cryo1,cryo2,cryo3,cryo5,cryo6}.
Such amplifiers, however, present significant challenges as careful selection of the active component and extra wiring for biasing are needed.
In addition, dissipation in the active component is often undesirable. 
In lack of detailed noise analysis, the conditions under which cryogenic TIAs are beneficial remain unknown.

We present a single stage wide band TIA which operates at room temperature and which
is designed for low noise measurements of QTFs in applications requiring long cables. 
We present a careful analysis of the frequency response and noise performance and show that
they remain virtually unaffected by the 180~pF cable capacitance of our setup.
For QTFs at room temperature we find that the amplifier noise is negligible as compared to the intrinsic thermal noise of the QTF.
In contrast, the amplifier noise prevails for QTFs at liquid helium temperatures. To our surprise, 
the dominant noise term is due to the voltage noise rather than the current noise of the TIA.
The overall noise performance of our TIA is either superior to or similar to that of cryogenic TIAs \cite{cryo1,cryo3,cryo5,cryo6}.

\begin{figure}[t]
{\centering
\includegraphics[width=.5\textwidth]{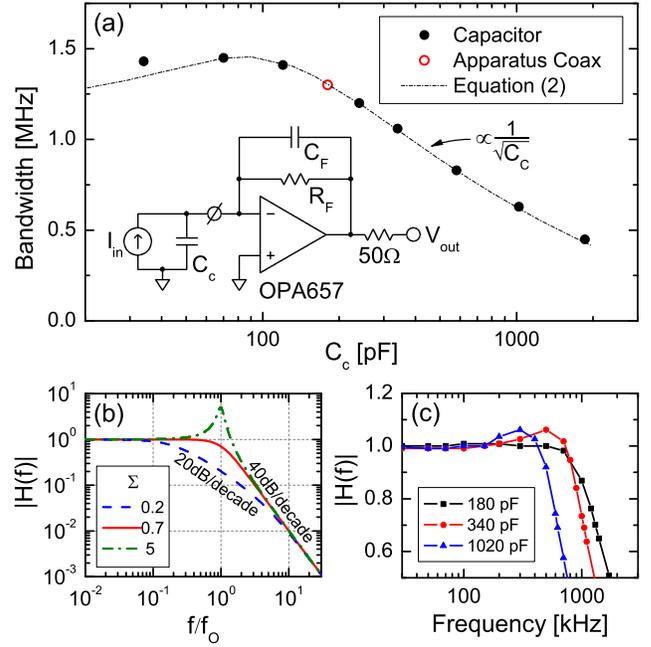}
}
\label{bandwidth}
\caption{ Amplifier bandwidth for several values of the input capacitance (panel a). Dots are measured values and the continuous line
is the prediction of Eq.2. The three qualitatively different frequency responses of TIAs (panel b). The measured frequency response of our TIA
for three different cable capacitances $C_C$ (panel c). }
\end{figure}

The circuit of our TIA is shown in Fig.1a. We make use of the low bias current OPA657 op-amp from Texas Instruments. 
The transimpedance gain is set by the $R_F=1$~M$\Omega$ feedback resistor. This resistor has a size 0805 surface mount packaging
which, together with the circuit board, adds a small parasitic capacitance $C_F\sim0.14$~pF. 
The capacitor at the input $C_C$ represents either a fixed capacitor used to characterize the circuit or
the 180~pF coaxial cable which is part of our cryostat.
The circuit also contains a 50$\Omega$ output resistor necessary to inhibit oscillations 
for capacitive loads and customary pairs of bypass capacitors of 10~nF and 10~$\mu$F.
Circuit components, with the exception of $R_F$ mounted on the back side of the circuit board, are visible in Fig.2a. 

The frequency response of TIAs is quite different from that of the more widely used voltage amplifiers.
The latter have a single-pole response similar to that of a low pass $RC$ filter. In contrast, TIAs have a more 
complicated response. Using a single-pole model $A(f)=A_{OL}f_A/(jf+f_A)$
for the op-amp, the magnitude of the dimensionless transfer function $|H(f)|$ for a 
feedback TIA, such as ours, is \cite{ti}
\begin{equation}
|H(f)|=\left| \frac{v_{out}}{R_F i_{in}} \right|=\frac{f_0^2}{\sqrt{(f^2-f_0^2)^2+f^2f_{0}^2  / \Sigma^2}}.
\end{equation}
Here $A_{OL}$ is the dc open loop gain and $f_{A}$ is corner frequency of the open loop gain of the op-amp. 
For large $A_{OL}$ and large cable capacitances $C_C \gg C_F$ the constants in the above equation relate to the 
circuit parameters as follows \cite{ti}
$f_0 = \sqrt{A_{OL} f_A/2 \pi R_F C_C}$ and $f_0/\Sigma = f_A(1+A_{OL}C_F/C_C)$.  
The bandwidth $f_{3dB}$ of the circuit can be expressed as
\begin{equation}
f_{3dB}^2 =f_0^2\left[\sqrt{(1-1/2 \Sigma^2)^2+1}+(1-1/2 \Sigma^2)\right].
\end{equation}

It is often unappreciated that the TIA transfer function has three qualitatively distinct behaviors.
As seen in Fig.1b, for $\Sigma>0.70$ there is a peak in the response at $f=f_0$ followed by a 40~dB/decade roll-off at higher frequencies.
This peaking gain may drive the circuit to saturation and it is commonly avoided by increasing $C_F$\cite{ti}.
For $\Sigma \ll 0.70$, the frequency response is flat up to $\Sigma f_0$, it has a 20~dB/decade roll-off for $f_0 \Sigma \lesssim f \lesssim f_0/\Sigma$,
and it has a 40~dB/decade roll-off at higher frequencies. It can be shown that in the 20~dB/decade region the circuit works as a charge amplifier \cite{uhv}.
Finally, for the limiting case of $\Sigma=0.70$ one obtains a maximally flat response followed by a 40~dB/decade roll-off
at higher frequencies. In this case the bandwidth can be approximated as\cite{ti} 
\begin{equation}
f_{3dB} \simeq \sqrt{A_{OL} f_A/2 \pi R_F C_C}.
\end{equation}
Here $A_{OL} f_A$ is the familiar gain bandwidth product $GBP$ of the op-amp and
the OPA657 was chosen because of the exceptionally high 1.6~GHz value of this product.

In Fig.1c we show the measured frequency response of our TIA for different $C_C$.
We find that for $C_C=180$~pF of our coax the frequency response is nearly maximally flat and the 
bandwidth is exceptionally large 1.3~MHz. This value agrees well with the prediction of Eq.3.
For $C_C$ up to 2~nF the bandwidth clearly remains wider than 32~kHz, the resonant frequency of the most common QTFs. 
In Fig.1a we plot the bandwidth extracted from our measurements and that expected from Eq.2. When using the parameters
$A_{OL}=5\times10^3$ and $f_A=0.32$~MHz of the OPA657, the two compare favorably for a wide range of $C_C$.
Furthermore, since for our TIA the parameter $\Sigma$ is close to 0.70 when $C_C \gtrsim 200$~pF, for this range of capacitances
Eq.3 predicts a bandwidth that scales approximately as $1/\sqrt{C_C}$. 

\begin{table*}[t]
\centering
\begin{tabular}{|c||c|c|c|c|c|c|c|c|}
\hline 
$T_{QTF}$~[K] & $R_{QTF}$~[k$\Omega$] & $\sqrt{4k_BT_{QTF}/R_{QTF}}$ & $e_{op}/R_{QTF}$ & $ 2\pi f e_{op} C_C$ & $\sqrt{4k_BT_F/R_F}$ & $e_{out}$~[$\mu$V/$\sqrt{\mbox{Hz}}$] \\ \hline
300 	& 17 	 	&  0.99 &  0.28	  &  0.18 	 	& 0.13 &  1.0 \\
4.2 	& 1.7   &  0.37	&  2.8    &  0.18     & 0.13 &  2.8 \\
\hline
\end{tabular}
\caption{ The parameters of the QTF, the contributions of different noise terms, and 
the prediction for the output noise  $e_{out}$ of the TIA. Noise estimations are done at the resonance frequency $f$ of the QTF
and using $C_C=180$~pF of our cable. Where not labeled, noise terms are measured in units of pA/$\sqrt{\mbox{Hz}}$.}
\label{table}
\end{table*}

QTFs are modeled as a series resonant LCR circuit connected in parallel with an additional capacitance \cite{coldq1}. 
However, near the resonance the approximation of a purely resistive impedance $R_{QTF}$ works well \cite{quartz:calibrate}.
Our TIA can be conveniently used to determine $R_{QTF}$ of the Citizen CFS308 QTF when using a 10~mV excitation. 
During the measurements the  QTF remained in its vacuum seal canister. The obtained $R_{QTF}$ values at 300~K and 4.2~K are listed in Table.I.

\begin{figure}[t]
{\centering
\includegraphics[width=.5\textwidth]{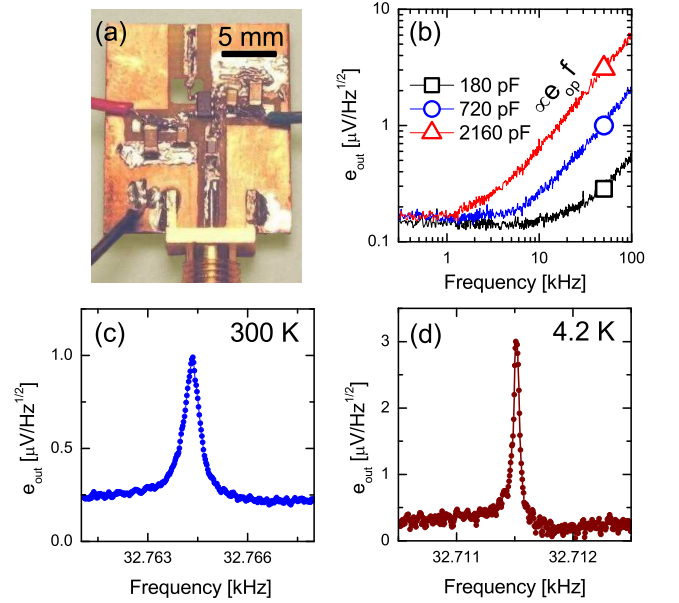}
}
\label{noisemodel}
\caption{ A photo of the circuit layout (panel a). The amplifier output noise at three different input capacitances $C_C$ (panel b).
The amplifier output noise at 300 K (panel c) and 4.2 K (panel d) when a QTF is connected to its input.  }
\end{figure} 

In the following we discuss the noise analysis of our TIA with a QTF connected to its input. 
In order to determine the dominant term in the power spectrum density $e_{out}$ measured at the output of the TIA,
it is useful to separate contributions due to the QTF and to the TIA itself. The former 
consists of the thermal noise $\sqrt{4 k_B T_{QTF}/R_{QTF}}$ at the resonant frequency of the QTF. 
We used $k_B$ for Boltzmann's constant and $T_{QTF}$ for the temperature of the QTF.
The amplifier noise $i_{amp}$ has four major contributions: one 
due to the current noise of the op-amp $i_{op}$, one due to the voltage noise of the op-amp $e_{op}$, 
one due to the presence of the capacitance $C_C$ at the input of the TIA,
and one due to the thermal fluctuations in the feedback resistor $R_F$. Since noises add in quadrature,
the amplifier noise can be expressed as\cite{ti}
\begin{equation}
\label{noise1}
i_{amp}^2= i_{op}^2 + (e_{op}/R_{QTF})^2 + (2\pi e_{op} f C_C)^2 + {4k_BT_F}/{R_F}.
\end{equation}
Adding to this the thermal noise of the QTF we find the noise at the output of the TIA is
\begin{equation}
\label{noise2}
e_{out}^2= R_F^2 \left( {4k_BT_{QTF}}/{R_{QTF}}+i_{amp}^2 \right).
\end{equation}
By establishing the dominant term in $e_{out}$ we can identify possibilities for lowering the noise levels and, furthermore, we may
compare the noise performance of various TIAs.

The noise parameters of the op-amp $i_{op}$ and $e_{op}$ may be read from the op-amp specification sheet. However, the
residues of the soldering flux left on the circuit board may cause excess noise.
To avoid this, we opened a hole in the circuit board near the inverting input which is visible in Fig.2a.
To estimate $e_{op}$ we measured $e_{out}$ with three different capacitances connected to the input
in the regime in which the frequency dependent third term in Eq.4 becomes significant. 
As seen in Fig.2b, this occurs at high frequencies and the
data is consistent with the $e_{op}=4.8$~nV/$\sqrt{\mbox{Hz}}$ specified for the OPA657.

With the measured $e_{op}$ we calculate the expected noise at the output $e_{out}$ at the resonant frequency of the QTF. 
The values at two different temperatures of our QTF are listed in Table.I. The extremely low $i_{op}=1.3$~fA/$\sqrt{\mbox{Hz}}$ 
of the OPA657 can be safely neglected. Measured noise data at 300~K and 4.2~K are shown in Fig.2c and Fig.2d,
respectively. We find that the measured peak values of $e_{out}$ are in good agreement with the estimations from Table.I and, hence,
our noise model is satisfactory.

We note that because of the LCR series circuit model of the QTF, at resonance $e_{out}$ exhibits a peak \cite{amp:limit} such as the ones in Fig.2.
However, such a peak could be either due to the thermal noise of the QTF or due to amplifier noise, depending on which one dominates.
By inspecting the noise estimations listed in Table.I, we find that when the QTF is at room temperature
the amplifier noise is negligible and, therefore, the peak in Fig.2c is due
to the thermal noise of the QTF \cite{uhv,amp:limit}. Any attempts to lower the noise of the TIA are unnecessary in this case. 
In contrast, when the QTF is at the temperature of liquid helium, the peak in $e_{out}$ shown in Fig.2d is dominated 
by the amplifier noise $i_{amp}$. Numbers from Table.I show that virtually all the noise comes from the $e_{op}/R_{QTF}$ term. 
We thus find that for low noise measurements of cold QTFs the voltage noise of the TIA rather than its current noise has to be minimized.
For our QTF the capacitance-dependent $2 \pi f e_{op} C_C$ term remains negligible both at 300~K and at 4.2~K.


We compare our room temperature TIA to cryogenic TIAs \cite{cryo1,cryo2,cryo3,cryo5,cryo6}. 
One TIA\cite{cryo1} has 5 times the noise for a similar $R_{QTF}$, hence we infer a voltage noise which is 5 times larger than ours.
Another TIA\cite{cryo5} has 26pA/$\sqrt{\mbox{Hz}}$ at 55~kHz which is about 80 times larger than ours at the same frequency.
The TIA in Ref.\cite{cryo6} has 100nV/$\sqrt{\mbox{Hz}}$ at room temperature and most likely remains much
larger than ours when cooled. The TIA which uses a JFET transistor as a first stage\cite{cryo3} has 3~nV/$\sqrt{\mbox{Hz}}$ which is 37\% smaller than ours.
In the absence of sufficient data, we are unable to compare our results to those in Ref.\cite{cryo2}.
We find, therefore, that there is very little or no improvement in the noise performance of cryogenic TIA as compared to ours.
For needle oscillators of high frequency \cite{app:extension1MHz},
however, $2 \pi f e_{op} C_C$ will be significant, hence important improvements result when reducing $C_C$.
Finally we note that a recently published room temperature TIA 
is expected to have a low voltage noise near 1~nV/$\sqrt{\mbox{Hz}}$ \cite{rt2}. 
This is, however, not a single stage but a discreet JFET/op-amp hybrid circuit and, therefore, it is considerably
more difficult to be built than ours.

To summarize, we built a low noise single stage TIA intended to be used for QTF measurements.
We have demonstrated a very large bandwidth of 1.3~MHz at 180~pF cable capacitance. We have provided a detailed noise analysis
and found that our amplifier adds negligible amount of noise when the QTF is at room temperature. For cold QTF, however,
the voltage noise of the op-amp becomes important and we provided a useful way to measure it. 
The noise due to the cable capacitance is found to be negligibly small. 
This work was supported by the DOE BES under contract no. DE-SC0006671.



\begin{thebibliography}{l}
\bibitem{karrai} K. Karrai and R.D. Grober, Appl. Phys. Lett. \textbf{66}, 1842 (1995).
\bibitem{intro:afm} H. Edwards {\it et al.}, J. Appl. Phys. \textbf{82}, 980 (1997).
\bibitem{intro:He3} R. Blaauwgeers {\it et al.}, J. Low Temp. Phys. \textbf{146}, 537 (2007).
\bibitem{intro:biosensor}X. Su {\it et al.}, Biosens. Bioelec. \textbf{17}, 111 (2002).
\bibitem{magn} M. Todorovic and S. Schultz, Appl. Phys. Lett.  \textbf{73}, 3595 (1998).
\bibitem{intro:thermometry} N. Samkharadze {\it et al.}, Rev. Sci. Instrum. \textbf{82}, 053902 (2011).

\bibitem{coldq1} J. Rychen {\it et al.}, Rev. Sci. Instrum. \textbf{70}, 2765 (1999).
\bibitem{coldq2} J. Rychen {\it et al.}, Rev. Sci. Instrum. \textbf{71}, 1695 (2000).
\bibitem{coldq3} R.H. Smit {\it et al.}, Rev. Sci. Instrum. \textbf{78}, 113705 (2007).
\bibitem{coldq4} J. Senzier {\it et al.}, Appl. Phys. Lett. \textbf{90}, 043114 (2007).
\bibitem{coldq5} A.E. Gildemeister {\it et al.}, Rev. Sci. Instrum. \textbf{78}, 013704 (2007).
\bibitem{coldq6} K. Saitoh {\it et al.}, J. Low Temp. Phys. \textbf{150}, 561 (2008).
\bibitem{coldq7} J.A. Hedberg {\it et al.}, Appl. Phys. Lett. \textbf{97}, 143107 (2010).
\bibitem{uhv} F.J. Giessibl {\it et al.}, Phys. Rev. B \textbf{84}, 125409 (2011).

\bibitem{heike} S. Heike and T. Hashizume, Appl. Phys. Lett. \textbf{83}, 3620 (2003).
\bibitem{app:extension650kHz}Z. Peng and P. West, Appl. Phys. Lett. \textbf{86}, 014107 (2005).
\bibitem{app:extension1MHz} T. An {\it et al.}, Rev. Sci. Instrum. \textbf{79}, 033703 (2008).


\bibitem{cryo1} C.H. Yang {\it et al.}, Rev. Sci. Instrum. \textbf{73}, 2713 (2002).
\bibitem{cryo2} N.G. Patil and J. Levy, Rev. Sci. Instrum. \textbf{73}, 486 (2002).
\bibitem{cryo3} K.R. Brown, L. Sun, and B.E. Kane, Rev. Sci. Instrum. \textbf{75}, 2029 (2004).
\bibitem{cryo5} D. Antionio {\it et al.}, Rev. Sci. Instrum. \textbf{79}, 084703 (2008).
\bibitem{cryo6} K. Hayashi {\it et al.}, J. Phys.:Conf. Ser. \textbf{150}, 012016 (2009).

\bibitem{ti} X. Ramus, Texas Instruments Application Report SBOA122.
\bibitem{quartz:calibrate} Y. Qin and R. Reifenberger, Rev. Sci. Instrum. \textbf{78}, 063704 (2007).
\bibitem{amp:limit} R.D. Grober {\it et al.}, Rev. Sci. Instrum. \textbf{71}, 2776 (2000).





\bibitem{rt2} Tzu-Yung Lin {\it et al.}, Rev. Sci. Instrum. \textbf{82}, 124101 (2011).

\end{thebibliography}
\end{document}